# Workflow patterns in process modeling

**Assist.Prof. Alexandra Fortiş, Ph.D. Candidate**
Universitatea „Tibiscus" din Timişoara
**Assist.Prof. Florin Fortiş, Ph.D.**
Universitatea de Vest Timişoara

ABSTRACT. This paper proposes an introduction to one of the newest modeling methods, an executable model based on workflows. We present the terminology for some basic workflow patterns, as described in [WfMC99].

## 1 Introduction

Through this paper, the authors are proposing a presentation of the notions that are creating the base of one of the most modern modeling methods: modeling through workflows. A workflow based model has to identify three aspects of the modeled process: the definition of the process, the process management (or its choreography), the effective course of the process (its orchestration).

The workflow model is an executable one, which can be read, executed and controlled by any workflow management system. It is a comprehensive model with extension possibilities in order to describe the process and to plan activities.

### 1.1 General notions

The goal of this paper is to present the terminology involved in the study through some definition exposed according to the specification imposed by the Workflow Management Coalition [WfMC99].

**Workflow Management Coalition** (WfMC) is a non-profit organization having as main objective the promotion of exploitation opportunities of technologies lied to workflow, through the development of a language and of some joint standards.

**Business Process Management Initiative** (BPMI) is a non-profit corporation which is trying to boost companies to develop and to operate business processes involving multiple applications and partners via Internet. Its main objective is the promotion and the development of some complete standards, XML-based, not affected by copyright. Such standards have to support and to allow the management of business processes in industry.

**Object Management Group** (OMG) is a non-profit consortium producing and maintaining specifications for the IT industry, for interoperable applications between enterprises.

**Definition 1.1** Through workflow one understands the *automation of a business process, in whole or in part, during which documents, information or tasks are passed from one participant to another for action, according to a set of procedural rules.*

**Definition 1.2** The Workflow Management System (WfMS) *is a system that defines, creates and manages the execution of workflows through the use of software, running on one or more workflow engines, which is able to interpret the process definition, to interact with workflow participants and, where required, to invoke the use of IT tools and applications.*

The workflow engine (WfE) is offering software services for the execution of an instance within a process

**Definition 1.3** A business (BP) process consists *in a set of one or more linked procedures and activities which collectively release a business policy goal, normally within the context of an organizational structure defining functional roles and relationships.*

**Definition 1.4** A process is a formal point-of-view on the business process, *supporting automated manipulation such as modeling or enactment by a workflow management system. It is a set of parallel or serial activities and their relationships, criteria to indicate the beginning and the termination of a process and information about individual activities, such as participants, associated IT applications and data.*

**Definition 1.5** *Through an activity we have a description for a work stage representing a logical step within a process.*

An activity included in a workflow requests human or automated resources in order to sustain the process execution. In the case of human resources, an activity is allocated to one of the workflow participants, through this action understanding a resource which executes the task representing the instance of a workflow activity.

**Definition 1.6** *An instance of an activity is the representation of an action during a single process execution.*

The instance of an activity or of a process, at a single execution, includes also some associated data. Each instance represents an individual execution thread of the process, it can be independently controlled, and it will have its own internal state as well as an external visible identity that can be used later on.

**Definition 1.7** *The Business Process Management Notation is a standard which creates a standardized path to fulfill the gap between the analysis and the implementation of a business process.*

The main objective for the introduction of this notation was to offer a common language easily understandable for all the participants in the process, from economic analysts which are initializing the process until the technology specialists which are responsible with the implementation of applications that will execute the process.
Another goal, equally important, is the insurance of the fact that the XML languages designed for the execution of the processes have to be viewed through a common notation.

**Definition 1.8** *YAWL - Yet Another Workflow Language – is a modeling language for workflows and it stands on the formalism of the Petri nets, enriched with patterns arising from the analysis of existent products for workflow modeling.*

## 2 Workflow elementary patterns

In this study we are going to introduce basic workflow patterns, as exposed in [A+04], [OMG03] and [WfMC99]

## 2.1 The Sequence Pattern

**Definition 2.1** *In a sequence, an activity of the process is enabled only after the completion of preceding activities and it will enable the following activity of the process.*

The sequence pattern is used to model consecutives stages in a workflow process. The main characteristic of this pattern is that of unconditioned interconnection between activities. Being an elementary pattern, it will be supported by any WfMS.

WfMC defines a similar notion: sequential routing, in order to model business processes [WfMC99], as follows:

**Definition 2.2** *A segment of a process instance under enactment by a workflow management system, in which several activities are executed in sequence under a single thread of execution, with no conditions between connected activities, is corresponding to sequential/serial routing.*

Graphical representations of this pattern in UML, YAWL or BPMN do not present notable differences, as can be seen in Figure 1. The plotted objects, independent of the modeling language, are representing activities and (sequential, unconditioned) relations between them.

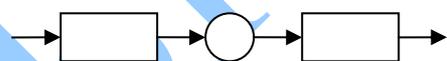

a) YAWL Representation

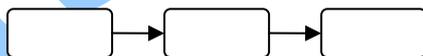

b) UML Representation

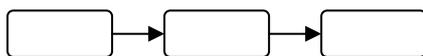

c) BPMN Representation

**Figure 1 Representation of the sequence pattern**

## 2.2 Parallel split pattern

**Definition 2.3** *In parallel split we are referring to a point in the workflow process where a single thread of control splits into multiple threads of control which can be executed in parallel.*

The parallel split pattern is used to model activities that can be executed in simultaneously or in any order. Again, being an elementary pattern it will be supported by any WfMS.

Two different approaches are available for this pattern, from the implementation point of view: *explicit* parallel split and *implicit* parallel split.

WfMC defines a similar notion for business process modeling, [WfMC99] – the parallel routing, as follows:

**Definition 2.4** *Through parallel routing for a segment of a process instance under enactment by a workflow management system, one understands that two or more activity instances are executing in parallel within the workflow, giving rise to multiple threads of control.*

For the YAWL representation we use the specific constructor for this pattern (AND-split) while for UML we use the fork operator for an explicit AND-split. In BPMN a specific operator is introduced for parallel routing but, in general, only the explicit parallel split is used, situation that does not require the presence of this operator.

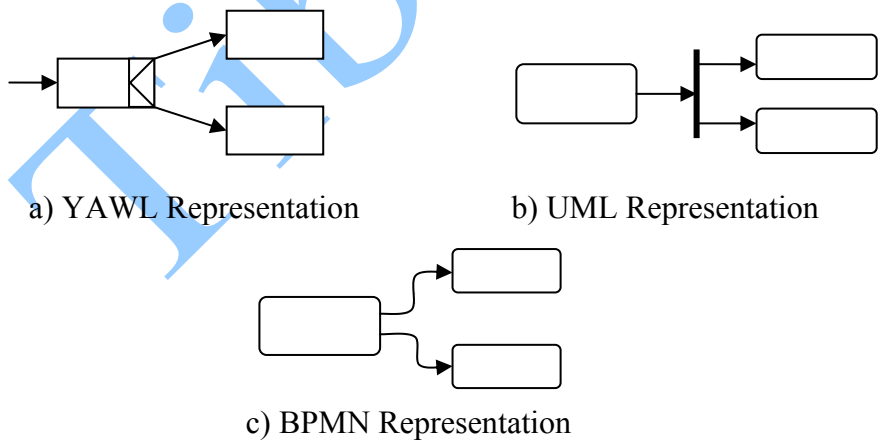

a) YAWL Representation        b) UML Representation

c) BPMN Representation

**Figure 2 Representation of the parallel split pattern**

## 2.3 Synchronization pattern

**Definition 2.5** *In the synchronization pattern we refer to a point in the workflow where multiple parallel subprocesses/activities converge into one single thread of control, thus synchronizing multiple threads.*

This pattern is used to model the convergence point of some parallel activities. An activity will be enabled only after the completion of all parallel converging synchronized activities. As elementary pattern it will be supported by any WfMS.

For this pattern we also have two different approaches for implementation: for the *explicit* synchronization and for implicit synchronization. For implementation we suppose that each incoming branch is executed only once. WfMC defines a similar notion, for business processes modeling [WfMC99] called the AND-join:

**Definition 2.6** *The AND-join is representing a point in the business process under enactment by a workflow management system in which two or more activities which have been simultaneously executed converge into a single control point. Each incoming branch executed in parallel is stopped until the set of all transitions to the next activity is completed which is equivalent to a moment when al branches are converging to a junction and the next activity is ready to be enabled.*

For this pattern, YAWL uses a specific operator called AND-join and UML uses the synchronization for an explicit AND-join action. The operator used in BPMN is identical to the one used for the previous pattern, that one being capable to treat simultaneously both parallel split and synchronization situations.

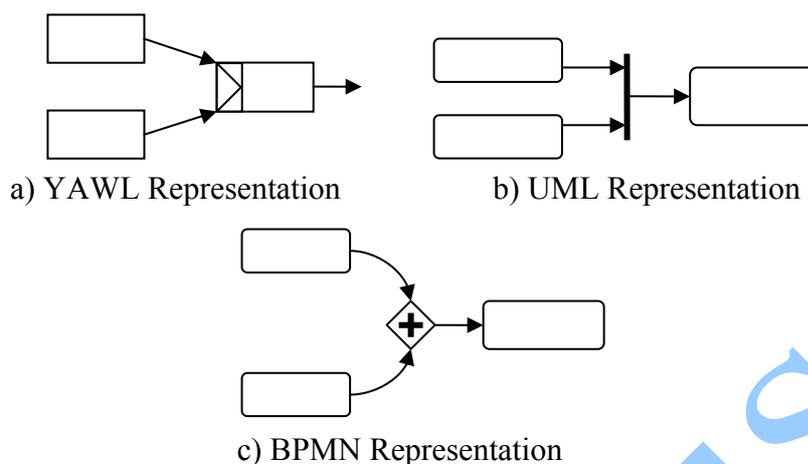

a) YAWL Representation  b) UML Representation

c) BPMN Representation

**Figure 3 Representation of the synchronization pattern**

## 2.4 Exclusive choice pattern

**Definition 2.7** *In the exclusive choice pattern we define a point in the workflow where, based on a data control decision, we choose exactly one of the multiple available outgoing branches.*

The exclusive choice pattern is use to model consecutive activities, with conditioned interconnection. Again, as an elementary pattern, it is supported by any WfMS.

For this pattern we have two available approaches for implementation: the *explicit* exclusive choice and *implicit* exclusive choice.

WfMC defines a similar notion called the exclusive split (XOR), for business processes modeling [WfMC99]:

**Definition 2.8** *The XOR split represents a point in a workflow modeling a business process, under enactment of a workflow management system, where a single thread of control makes a decision upon which branch to take when multiple alternative branches are available. The XOR split is a conditioned one and the transition to another activity is made according to the consequences of transition's conditions.*

In order to represent this pattern in YAWL, we use a specific operator called XOR-split and UML uses a choice operator for an explicit XOR situation. In BPMN we also have a specific operator to model this situation - the exclusive decision. This type of operator is implicitly used in decision situations.

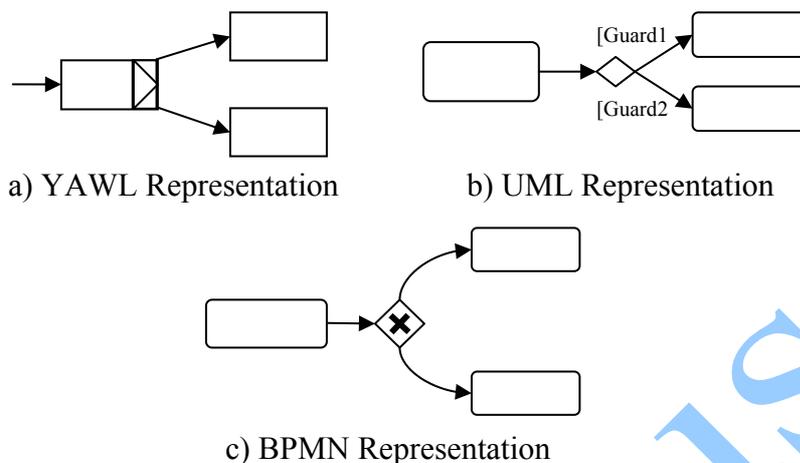

a) YAWL Representation    b) UML Representation

c) BPMN Representation

**Figure 4  Representation of the exclusive choice pattern**

## 2.5  Simple merge pattern

**Definition 2.9** *In the simple merge pattern, we define a point in the workflow where two or more alternative branches are joined, without synchronization. A conjecture feature for this pattern is that none of those alternative branches will be executed simultaneously.*

The pattern of simple merge appears in modeling consecutive activities in which an activity is enabled only when all of the precedent activities are completed. Further more, there is no parallelism relation before the current activity. As elementary pattern, the simple merge is supported by any WfMS.

WfMC defines a similar notion, the OR-join, to model business processes, [WfMC99] according to:

**Definition 2.10** *The OR-join is a point in a workflow modeling a business process, under the enactment of a WfMS, where two or more activity branches re-converge to a single common activity as the next step within the workflow. As no parallel activity execution has occurred at the join point, no synchronization is required.*

In YAWL we use a specific operator - XOR-join – while UML uses a join operator. BPMN again uses the exclusive choice pattern under one of the forms: XOR gate or exclusive decision gate, in order to solve this pattern modeling. However, in BPMN is possible to direct model the simple merge pattern, without using an alternative mechanism, but in this case the join operation can not be controlled.

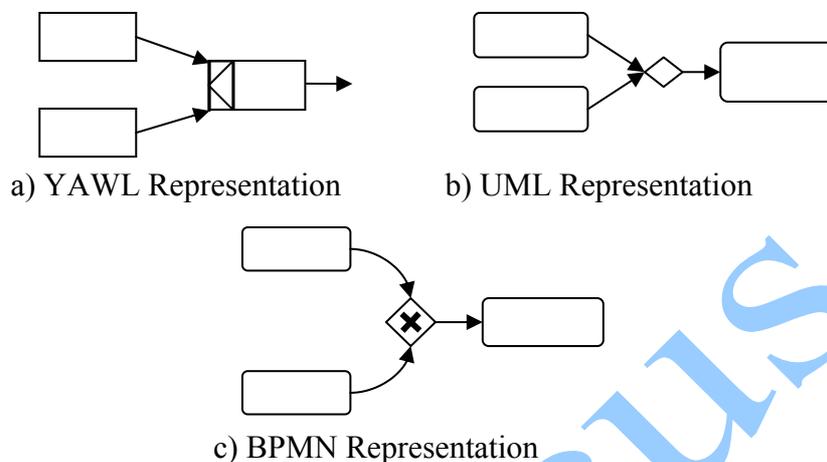

a) YAWL Representation   b) UML Representation

c) BPMN Representation

**Figure 5 Representation of the simple merge pattern**

## 3 Advanced workflow patterns

This section is dedicated to a presentation of advanced basic patterns, according to the work in the papers [A+04], [OMG03] and [WfMC99] patterns involved in the construction of workflows: multiple choices, synchronization, multiple join and discriminator.

### 3.1 The Multiple Choice Pattern

**Definition 3.1** *Through the multiple choice pattern, we define a point in the workflow process where, based on a decision or workflow control data, a number of branches are chosen.*

The pattern is used to model consecutive conditioned activities in which the transition through the next activity is made according to the conditions in the transition.

Most WfMS can directly integrate this pattern by specifying the conditions for the transitions and those that can not directly implement it use combination between the simple merge pattern and the exclusive choice pattern.

WfMC defines XOR-split and OR-split in the same manner. A difference in representation can be noticed in the case of UML.

89

For BPMN a new operator is introduced: an inclusive decision gate. As an alternative, BPMN offers the possibility to use some mini-gates of decision, directly linked with the action which precedes the moment of the multiple choice.

YAWL uses a special symbol, the OR-split, to model this pattern. A similar operator is defined also in BPMN, while UML uses, again, a fork operator to manage this situation.

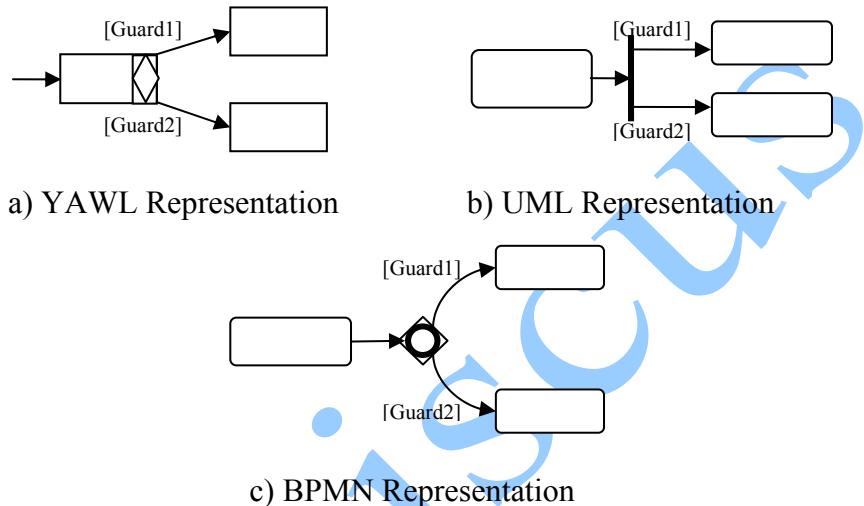

a) YAWL Representation        b) UML Representation

c) BPMN Representation

**Figure 6 Representation of the multiple choice pattern**

## 3.2 The Synchronizing Merge Pattern

**Definition 3.2** *Through the synchronizing merge pattern, we describe a point in the workflow where multiple paths from the flow converge to a unique thread. If one chooses more than a path, synchronization for the active branches is needed. For a single path choice, alternative branches converge through this point, without synchronization.*

The conjecture that characterizes this pattern is that, once enabled, a branch can not be re-enabled as long as the process is waiting for the completion of other branches.

The synchronizing merge pattern is modeling consecutive activities in a workflow in which the execution of the current activity is started if at least one of the previous activities is completed and there is no possibility that one of these parallel branches becomes active given the current state of the workflow.

One of the major problems arise from the implementation of this pattern is the decision making in order to establish if we are talking about synchronization or reunion. This type of reunion has to establish if it is necessary to wait for activation from one of the previous branches.

Most WfMS integrate this pattern through some combinations AND- and XOR-type, in order to avoid the simultaneous activation of more transitions. The synchronization of parallel branches is made through the AND-join and some simple join specific constructions.

YAWL uses a special symbol, OR-join, to model the synchronizing merge pattern, similar to an OR-split. A similar operator is defined by BPMN, while the UML solution is not so simple to obtain; White [Whi04] proposes a solution as in Figure 7b), but this solution catches only a part of the problem (see [W+04] for details).

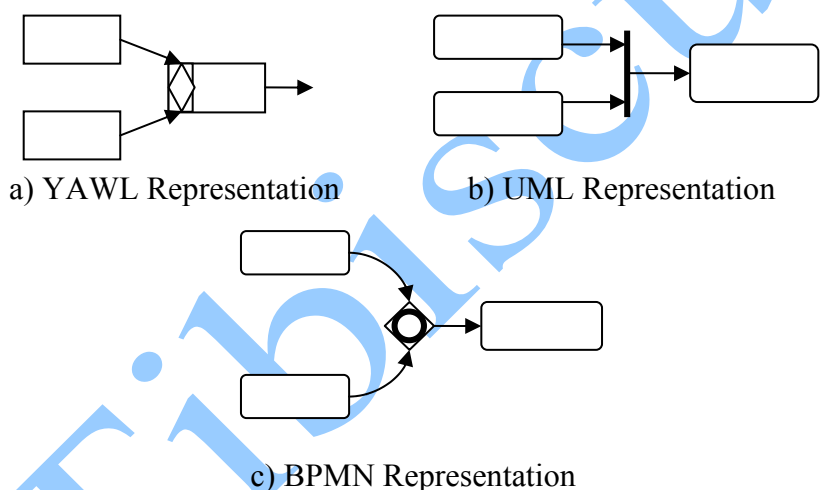

a) YAWL Representation      b) UML Representation

c) BPMN Representation

**Figure 7 Representation of the synchronizing merge pattern**

## 3.3 Multiple Merge Pattern

**Definition 3.3** *The multiple merge pattern denominates a point in the workflow in which two or more branches converge, without synchronization. If more than a branch is enabled, possible through concurrent processes, the activity following the reunion is launched at any activation of any branch preceding the activity.*

This pattern is used to model processes that are including two or more converging branches, through the same final node. For the multiple merge, WfMC defines the concept of OR-join. The representation of the multiple merge pattern is not modified related to the simple merge one.

For BPMN, this pattern does not make assumption on the existence of an additional operator. Possible decision gates are inefficient in this case, so this pattern will rely only on the BPMN capacity to allow multiple branches to penetrate an activity.

### 3.4 The Discriminator Pattern

**Definition 3.4** The *discriminator pattern defines a point in the workflow that waits for one of the incoming branches to complete before activating the subsequent activity. From that moment on it waits for all remaining branches to complete and "ignores" them. Once all incoming branches have been triggered, it resets itself so that it can be triggered again (which is important otherwise it could not really be used in the context of a loop).*

For this pattern implementation, two possibilities are available. The first situation deals with the fact that the first workflow branch that is completed will enable the next task, action that will cancel all the other branches. The second possibility stipulates that the next activity may begin only after the completion of a default number of branches in the workflow, all the other branches being ignored.

The discriminator is a special case of the „n-out-of-m" pattern, matching here a situation like in „1-out-of-m".

The solution offered by the discriminator pattern is not as simple to obtain as for the other patterns. In this sense, UML offers a solution which is, at the same time, a generalization of the pattern „n-out-of-m". The YAWL solution is similar with the UML solution, while BPMN uses an exclusive gate (XOR) to model this pattern. The difference in the use of this pattern arises from the presence of a possible uncontrolled parallel decomposition executed before the discrimination operation.

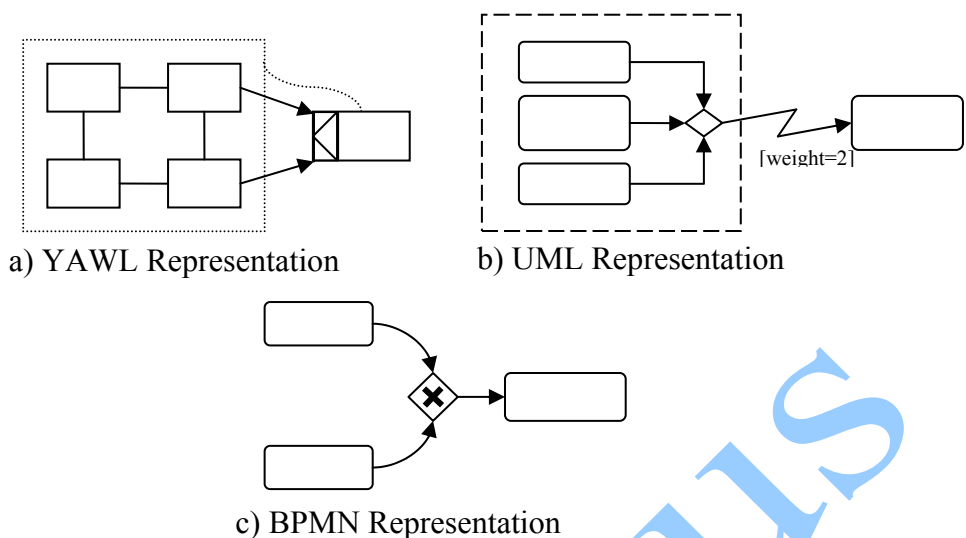

a) YAWL Representation    b) UML Representation

c) BPMN Representation

**Figure 8 Representation of the discriminator pattern**